# India's rank and global share in scientific research: *how publication counting method and subject selection can vary the outcomes?*

Vivek Kumar Singh[1*], Parveen Arora[2], Ashraf Uddin[3] & Sujit Bhattacharya[4]

[1]Department of Computer Science, Banaras Hindu University, Varanasi-221 005, India.
[2]Department of Science and Technology, Govt of India, New Delhi-110 016, India.
[3]Department of Computer Science, American International University-Bangladesh, Dhaka 1229, Bangladesh.
[4]CSIR- National Institute of Science Technology And Development Studies, New Delhi-110 012, India.

**Abstract:** During the last two decades, India has emerged as a major knowledge producer in the world, however different reports put it at different ranks, varying from 3[rd] to 9[th] places. The recent commissioned study reports of Department of Science and Technology (DST) done by Elsevier and Clarivate Analytics, rank India at 5[th] and 9[th] places, respectively. On the other hand, an independent report by National Science Foundation (NSF) of United States (US), ranks India at 3[rd] place on research output in Science and Engineering area. Interestingly, both, the Elsevier and the NSF reports use Scopus data, and yet surprisingly their outcomes are different. This article, therefore, attempts to investigate as to how the use of same database can still produce different outcomes, due to differences in methodological approaches. The publication counting method used and the subject selection approach are the two main exogenous factors identified to cause these variations. The implications of the analytical outcomes are discussed with special focus on policy perspectives.

**Keywords:** Fractional Counting, Indian research, Research output ranking, Scholarly databases, Whole counting.

## Introduction

In the present era of knowledge-based economy, countries that produce new scientific knowledge do well in economic development and are able to achieve prosperity and well-being of its citizens. The United Nation's 2030 Sustainable Development Goals also include several goals that can only be achieved through creation and application of new knowledge and technologies. It is because of these reasons that countries are investing more on Research & Development (R&D) activities. Given the increased focus on R&D by countries, several exercises are now caried out across the world for measuring research productivity at different levels. Many international as well as national reports are published focusing on assessment of research productivity, quality and impact. However, different reports often produce different outcomes with respect to relative positions of different countries. These varied outcomes often create confusion. For example, in the case of India, the most recent report on Research and Development Statistics[1] released by Department of Science and Technology (DST) of Government of India (on page 13), shows three different curves for India's research output rank, one ranking India at 3[rd], second at 5[th] and third at 9[th] place. The three curves apparently are drawn from three different reports. Therefore, it is important that the different factors shaping the outcomes of these assessment and ranking are understood well.

The National Science Foundation (NSF) report on Science and Engineering indicators[2] shows India ranked at 3[rd] place in global research output. The reports of the two commissioned studies of DST, done by Clarivate Analytics[3] (owner of Web of Science database) and by Elsevier[4] (owner of Scopus database), show India's research output rank as 9[th] and 5[th], respectively. DST`s Elsevier and Clarivate Analytics reports use data from different databases, namely Scopus and Web of Science, respectively, and therefore, one may understand that varied coverage of databases may be responsible for the different outcomes of these reports, as explained in detail in the first part of this study. (Singh et al., 2020)[5] This previous study showed how the endogenous factors (related to use of different databases) cause variations in findings of different reports. However, given that the NSF and DST- Elsevier reports, both are based on Scopus database, and yet produce different outcomes; it is imperative that exogenous factors (related to methodological approach used) also play an important role in shaping up the outcomes. Two key exogenous factors are identified to be responsible for the variations- the publication counting method (*whole counting* or *fractional counting*) used and the subject selection.

---

[*] Corresponding author. Email: vivek@bhu.ac.in

The article analyses the findings of the NSF Science and Engineering Indicators report and the DST-Elsevier report, and also some independently obtained data from Scopus database. The analytical results show that changing the publication counting method from *whole* counting to *fractional* counting significantly changes the outcome. Similarly, using data for different set (or subset) of subjects is also found to produce different evidence of research output volume rank of different countries. This article (along with the previous study[5]), therefore, tries to analyse the factors shaping the research output ranks of different countries and identify the reasons why different reports may produce different outcomes.

**Objectives**

The article attempts to identify the impact of exogenous factors, mainly publication counting method and subject selection, on the outcomes of research assessment exercises. Data and outcomes in the two reports (NSF and DST- Elsevier reports) as well as independently obtained data from Scopus database, are analysed for the purpose.

**Data and methods**

The study uses data from three sources: (a) NSF Science and Engineering Indicators Report, 2020, (b) DST- Elsevier commissioned study report 2019, and (c) Scopus database.

The NSF Science and Engineering (S&E) Indicators report obtained data[6] from Scopus database for the period 2000-2018 for S&E in 14 subject areas (S1) pertaining to Science and Engineering area. These subject areas include Agricultural Sciences, Astronomy & Astrophysics, Biological & Biomedical Sciences, Chemistry, Computer & Information Sciences, Engineering, "Geosciences, Atmospheric &Ocean Sciences", Health Sciences, Materials Science, Mathematics & Statistics, Natural Resources & Conservation, Physics, Psychology, Social Sciences. The data included in the analysis was for the document types of research article, review and conference paper (D1).

The DST- Elsevier report has also drawn data from Scopus database for a set of 16 core S&T subject areas (S2) during 2011-2016. These include Engineering, Medicine, Materials Science, Chemistry, Mathematics, Chemical Engineering, Energy, Immunology & Microbiology, Computer Science, Physics & Astronomy, "Biochemistry, Genetics, & Molecular Biology", "Pharmacology, Toxicology, & Pharmaceutics", Agriculture & Biological Sciences, Environmental Science, Earth & Planetary Sciences, Veterinary. The data included in the analysis was for the document types of research article, review and conference paper (D2).

We have mainly analysed the data for publication year 2016, mainly because (a) DST- Elsevier report only had data up to publication year 2016 and (b) publication year 2016 is one of the most recent periods with stable data. It may be noted that the two reports are just slightly different in their subject area selection, with NSF focused mainly on Science and Engineering (including some Social Sciences), whereas DST- Elsevier study included research output in all major areas of Science, Technology and Medicine. Thus, the two report mainly differ in coverage of subject areas like Social Science, Psychology, Health Science and Nursing.

In addition to analysing the data from the two reports, we have also independently obtained research output data for some selected subject groups from Scopus for 20 most productive countries for the publication year 2016. These groups (S3) include the four broad subject areas of Scopus: (a) Life Sciences, (b) Physical Sciences, (c) Health Sciences, and (d) Social Sciences and Humanities; and some selected subject areas:(e) Computer Science, (f) Social Science and Arts & Humanities, (g) Engineering, (h) Agriculture & Veterinary Science and Biology (i) Medicine, Pharmacology, Immunology, Health& Dental Science. The data was obtained for document types 'article', 'review', and 'conference paper' (D3). This independently obtained data was mainly used to analyse and show what impact the subject selection may have on research output rank of countries. It would be relevant here to mention that Scopus database uses a source-based subject classification, wherein articles are assigned to one or more subject areas based on the journal in which it is published. The journals are permanently assigned to selected subject areas.

The method for analytical study comprised of quantitative and computational approach. The results are shown in tables and figures, drawn mainly by using Excel functions and utilities.

*First of all*, the 2016 publication data from the NSF report for 20 countries was analysed by varying the publication counting methods, from *whole counting (WC)* to *fractional counting* (FC). For this purpose, the proportion of collaborative output of the countries was obtained from the report. The variations in scores due to use of the two counting methods are observed and analysed. The DST- Elsevier report is also analysed along with its outcomes for the publication year 2016. The research output ranks of different countries for *whole* counting of NSF and DST- Elsevier reports are compared and correlated.

*Secondly*, the relationship between the reduction in publication score due to *fractional* counting and internationally collaborated paper (ICP) percentage for different countries is analysed. The objective was to observe whether countries that engage in higher international collaboration stand to lose in publication score due to use of *fractional* counting.

*Thirdly*, the independently obtained research publication data for publication year 2016 for different countries for different subject areas was analysed. The data for different subject areas are compared to observe the quantum of variations in research output ranks across different subject areas. The Spearman Rank Correlations are also computed for different subject area-based research output ranks.

**Results**

The analytical results are organized into two parts. *First*, the impact of changing the publication counting method from *whole* to *fractional* on the research assessment outcome is analysed. The relationship between international collaboration and *fractional* counting is also presented. *Secondly*, the variations in research output volume and rank of different countries for different subject areas is observed and analysed.

### Impact of fractional vs. whole publication counting methods

The difference between *whole* and *fractional* counting methods can be understood from the fact that *whole* counting method gives equal score/ credit (score of 1) to each author (and hence the affiliating institution/ country), for each publication record. However, *fractional* counting method divides the score/ credit for each publication record among the authors (and hence the affiliating institution/ country). Thus, if there are more authors in a research article, the score that each author gets, would be equally divided among them. This implies that authors (and consequently affiliating institutions or countries) get lesser aggregate score of research publications if they publish more collaborated output involving higher number of authors.

For the 2016 data from NSF report, the research output ranks of 20 countries, using both the *whole* counting and *fractional* counting methods are computed. Table 1 shows the *whole* count and *fractional* count scores and research output ranks for the NSF report data, *whole* count for DST- Elsevier report data, and the rank correlations between the different ranks in them.

It is observed that countries like UK, France, Canada and Australia that have higher *whole* count score of publications get lower *fractional* count score, and consequently the research output rank. On the other hand, countries like India, Russia, South Korea and Brazil get higher *fractional* count score and rank despite having relatively lesser *whole* count of publications. Looking at data for some specific countries help understanding the impact of the counting method further. For example, we observe that, UK has 161,910 absolute number of publications and 3$^{rd}$ rank as per whole counting method. However, if *fractional* counting is used instead of *whole* counting, its publication score decreases to 97,680.90 and rank decreases to 6$^{th}$. A similar pattern is observed in cases of countries like France, Canada, Australia etc., which all stand to lose in terms of research output rank, while changing the counting method from *whole* to *fractional*.

A counter example is India, which has 150,013 absolute number of publications and 5$^{th}$ rank as per *whole* counting method. However, its score with *fractional* counting reduces to 1,35,787.79 but rank improves to 3$^{rd}$. Thus, India ranking 5$^{th}$ in terms of absolute number of publications moves to 3$^{rd}$ rank if *fractional* counting method is used. This indicates that the reduction in score for India due to *fractional* counting is lesser as compared to other countries like UK and Germany, both of which have higher absolute research output than India. Similar patterns are observed in case of countries like Russia, South Korea, Brazil etc. that stand to gain in terms of rank, while changing counting method from *whole* to *fractional*.

Thus, it is observed that countries which have higher collaborated output stand to lose more in case the *fractional* counting method is used. On the contrary, countries which have less collaborated output stand to gain in rank if *fractional* counting is used.

**Table 1: Publication count and ranks of different countries as per NSF & Elsevier report data for 2016 along with Spearman Rank Correlation Coefficient (SRCC) values**

| Country | NSF Report Data | | | | Elsevier Report Data | | Rank Correlation | |
|---|---|---|---|---|---|---|---|---|
| | 1 | 2 | 3 | 4 | 5 | 6 | 7 | 8 |
| | Whole Count (WC) | Rank WC | Fractional Count (FC) | Rank FC | Whole Count (WC) | Rank WC | SRCC (1,5) | SRCC (2,4) |
| USA | 541080 | 1 | 427264.63 | 2 | 546,548 | 1 | | |
| China | 483862 | 2 | 438348.74 | 1 | 469,441 | 2 | | |
| UK | 156899 | 3 | 99366.17 | 6 | 162,005 | 3 | | |
| Germany | 154913 | 4 | 108295.59 | 4 | 154,809 | 4 | | |
| India | 123977 | 5 | 112167.34 | 3 | 136,238 | 5 | | |
| Japan | 120505 | 6 | 101297.3 | 5 | 115,541 | 6 | | |
| France | 106846 | 7 | 71028.47 | 7 | 106,557 | 7 | | |
| Italy | 96822 | 8 | 70534.27 | 8 | 97,665 | 8 | | |
| Canada | 89219 | 9 | 60045 | 11 | 89,806 | 9 | | |
| Australia | 80404 | 10 | 53781.62 | 14 | 81,862 | 10 | 0.999 | 0.947 |
| Spain | 78642 | 11 | 55514.33 | 12 | 80,253 | 11 | | |
| South Korea | 74018 | 12 | 62735.09 | 9 | 77,215 | 12 | | |
| Russia | 73093 | 13 | 62661.74 | 10 | 75,595 | 13 | | |
| Brazil | 66813 | 14 | 55181.31 | 13 | 67,012 | 14 | | |
| Netherlands | 50971 | 15 | 31014.65 | 18 | 51,619 | 15 | | |
| Iran | 47548 | 16 | 42855.86 | 15 | 49,500 | 16 | | |
| Poland | 43087 | 17 | 34838.68 | 17 | 41,806 | 17 | | |
| Turkey | 41005 | 18 | 35510.17 | 16 | 41,405 | 18 | | |
| Switzerland | 40285 | 19 | 21952.33 | 19 | 39,938 | 19 | | |
| Sweden (Taiwan) | 35945 (34561) | 20 (21) | 20860.65 | 20 | Taiwan (34,770) | 20 | | |

In order to understand the differences in different ranks of NSF and DST- Elsevier reports, we have computed the Spearman Rank Correlation Coefficient (SRCC). The SRCC value between whole count rank of NSF and Elsevier is found to be 0.999. Thus, the two reports agree significantly in their research output ranks of countries (the first 19 ranks are actually similar, with difference in 20[th] place, with Taiwan in DST- Elsevier report and Sweden in NSF report). The SRCC value between *whole* and *fractional* count ranks of NSF report is found to be 0.947, which also indicate good agreement. However, the level of agreement is lesser than between *whole* counting ranks of NSF and DST- Elsevier reports. Thus, it can be observed that use of *fractional* instead of *whole* counting will cause more variations in ranks of countries that are close in publication volume but differ in their collaborated paper volumes. USA and China (with USA having at least 60,000 publications more than China) constitute an interesting example, as their relative research output ranks get affected with the use of *fractional* counting due to higher differences in their collaborated papers. UK is another example, which goes down to 6th rank with the use of *fractional* counting due to higher amount of collaborated papers. We will see below that the reduction in publication scores due to use of *fractional* counting are found to be highly correlated with the ICP instances of the countries.

We have analysed the relationship between (a) reduction in score due to use of *fractional* counting, and (b) internationally collaborated paper (ICP) percentage of different countries. The motivation was to see if those countries which have higher proportion of their research output as internationally collaborated, actually suffer in score due to use of *fractional* counting method. For this purpose, the proportion of ICP instances for all the 20 countries was obtained and correlated with reduction percentage. Table 2 shows the absolute number of publications, publication score reduction due to fractional counting, and ICP instances and percentage for the 20 countries. It is observed that countries like UK (61.7% ICP), France (58.4% ICP) and Australia (59.5% ICP) get much higher reduction in score as compared to other countries. Countries with lower ICP% are the ones to get lowest score reduction due to *fractional* counting. The last column of the table 2 shows the ratio of percentages of reduction and ICP. A higher value indicates higher loss of score of a country, connected to higher ICP instances. A ratio of greater than '1' for some countries is in a sense indication of intense multi-institutional, multi-country, and multi-disciplinary research collaboration, which unfortunately gets neglected with the use of *fractional* counting.

**Table 2: NSF report data for year 2016- publication scores, reduction due to FC and ICP values**

| Country | Whole Count (WC) | Fractional Count (FC) | Reduction in score due to FC (in terms of % of whole data) | Internationally Collaborated Papers (ICP) | ICP as % of whole data (1) | Ratio (3/5) |
|---|---|---|---|---|---|---|
| | 1 | 2 | 3 | 4 | 5 | 6 |
| USA | 541,080 | 427,264.63 | 26.6 | 198,875 | 36.8 | 0.72 |
| China | 483,862 | 438,348.74 | 10.4 | 98,327 | 20.3 | 0.51 |
| UK | 156,899 | 99,366.17 | 57.9 | 90,497 | 57.7 | 1 |
| Germany | 154,913 | 108,295.59 | 43 | 78,223 | 50.5 | 0.85 |
| India | 123,977 | 112,167.34 | 10.5 | 21,815 | 17.6 | 0.6 |
| Japan | 120,505 | 101,297.3 | 19 | 33,217 | 27.6 | 0.69 |
| France | 106,846 | 71,028.47 | 50.4 | 58,878 | 55.1 | 0.91 |
| Italy | 96,822 | 70,534.27 | 37.3 | 46,064 | 47.6 | 0.78 |
| Canada | 89,219 | 60,045 | 48.6 | 47,015 | 52.7 | 0.92 |
| Australia | 80,404 | 53,781.62 | 49.5 | 43,702 | 54.4 | 0.91 |
| Spain | 78,642 | 55,514.33 | 41.7 | 39,412 | 50.1 | 0.83 |
| South Korea | 74,018 | 62,735.09 | 18 | 20,560 | 27.8 | 0.65 |
| Russia | 73,093 | 62,661.74 | 16.6 | 18,032 | 24.7 | 0.67 |
| Brazil | 66,813 | 55,181.31 | 21.1 | 21,673 | 32.4 | 0.65 |
| Netherlands | 50,971 | 31,014.65 | 64.3 | 31,414 | 61.6 | 1.04 |
| Iran | 47,548 | 42,855.86 | 10.9 | 9,794 | 20.6 | 0.53 |
| Poland | 43,087 | 34,838.68 | 23.7 | 13,218 | 30.7 | 0.77 |
| Turkey | 41,005 | 35,510.17 | 15.5 | 9,112 | 22.2 | 0.7 |
| Switzerland | 40,285 | 21,952.33 | 83.5 | 27,791 | 69 | 1.21 |
| Sweden | 35,945 | 20,860.65 | 72.3 | 22,929 | 63.8 | 1.13 |

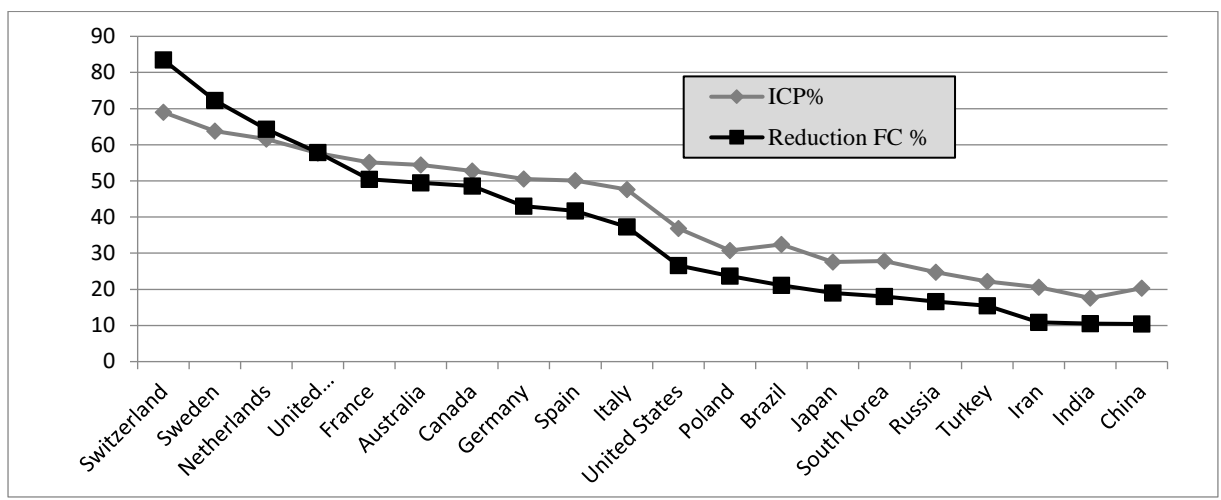

**Fig. 1-- ICP% vs. Reduction% due to Fractional Count for the 20 countries [Data Source: NSF Report 2020]**

To illustrate the relationship between reduction percentage and ICP % further, Fig. 1 presents a plot of ICP% and reduction%, ordered in descending order of the values for the 20 countries. It is observed that these two curves correlate well, with a Pearson Correlation Coefficient value of 0.98. This implies that ICP% and reduction% are strongly related, with higher ICP% indicating higher reduction% of score due to *fractional* counting. Therefore, it can be understood that there is a definite relationship between ICP and *fractional* counting method. The use of *fractional* counting method is observed to reduce the publication score of countries that have higher proportion of internationally collaborated output. Thus, though the use of *fractional* counting may be suggested[7,8] for research performance assessment at different levels of granularity, it is observed that using it for country-level studies results in masking of the important dimension of research collaboration networks of countries.

**Analysing the impact of subject selection**

The next exogenous factor that we analysed is the subject selection in different reports and its impact on the outcomes. The data downloaded independently from Scopus for 20 most productive countries for the publication year 2016 for document types 'article', 'review', and 'conference paper' for different subject areas is analysed for the purpose. The research output data for the four major areas of Scopus as well as some selected specific subjects, as described in the data section, are analysed and the research output ranks are computed. Table 3 shows the research output volume and ranks of the 20 most productive countries.

It is observed that research output ranks of the 20 countries vary across both, the major and the specific subjects. For example, if we look at major areas, it is observed that India is ranked at 3rd rank in research output in Physical Sciences, 5th in Life Sciences, and 10th in Health Sciences and Social Science, both. It may be seen that the four most productive countries- USA, China, UK and Germany are more or less at same research output rank in all the four major areas. The variations in relative ranks across different major areas are seen more in case of some countries, lower in the rank. One example is Russia, which is at 8th rank in research output in Physical Sciences, 12th in Social Sciences, 17th in Life Sciences and 23rd in Health Sciences. In terms of research output in all fields taken together, Russia is at 12th rank.

Similar variations are also seen for research output in specific subjects. For example, USA is at 1st rank in overall data but at 2nd in CS research output. China is at 2nd rank in overall data but at 1st rank in CS research output. UK, which is at 3rd rank on overall data, is at 4th rank in CS, 2nd in SS&AH, 4th in ENG and AGR, BIO & VET, and 3rd rank in MED, IMM & DEN. India also has variation in ranks, with rank 5th in overall, 3rd in CS and 10th in MED, IMM & DEN. Another interesting case is Japan, which is ranked 6th overall, but 11th in CS, 21st in SS&AH, 7th in ENG, 12th in AGR, BIO & VET, and 5th in MED, IMM & DEN. Thus, the relative research output ranks of different countries vary significantly across different subject areas.

**Table 3: Research output and rank of different countries in various subject areas for the year 2016 as per Scopus data**

| Country | ALL Fields | | Health Sciences | | Life Sciences | | Physical Sciences | | Social Science | | CS | | SS & AH | | ENG. | | AGR, BIO, VET | | MED, IMM & DEN | |
|---|---|---|---|---|---|---|---|---|---|---|---|---|---|---|---|---|---|---|---|---|
| | 1 | 2 | 3 | 4 | 5 | 6 | 7 | 8 | 9 | 10 | 11 | 12 | 13 | 14 | 15 | 16 | 17 | 18 | 19 | 20 |
| | TP | R | TP | R | TP | R | TP | R | TP | R | TP | R | TP | R | TP | R | TP | R | TP | R |
| USA | 574705 | 1 | 202087 | 1 | 146895 | 1 | 258315 | 2 | 105178 | 1 | 62875 | 2 | 72433 | 1 | 102951 | 2 | 47685 | 1 | 201151 | 1 |
| China | 487196 | 2 | 84881 | 2 | 108614 | 2 | 355630 | 1 | 23545 | 3 | 77341 | 1 | 11749 | 5 | 187942 | 1 | 36031 | 2 | 86346 | 2 |
| UK | 170925 | 3 | 57345 | 3 | 39042 | 3 | 76659 | 5 | 37672 | 2 | 18711 | 5 | 26168 | 2 | 28748 | 6 | 13523 | 4 | 55396 | 3 |
| Germany | 161325 | 4 | 45471 | 4 | 38557 | 4 | 90793 | 4 | 20936 | 4 | 20968 | 4 | 12255 | 3 | 33744 | 4 | 12951 | 5 | 44827 | 4 |
| India | 142562 | 5 | 28305 | 10 | 35331 | 5 | 93821 | 3 | 11021 | 10 | 30131 | 3 | 5979 | 12 | 42511 | 3 | 11464 | 6 | 34491 | 6 |
| Japan | 121416 | 6 | 37592 | 5 | 30450 | 6 | 70788 | 6 | 6041 | 15 | 16579 | 6 | 3474 | 18 | 32095 | 5 | 9182 | 11 | 37908 | 5 |
| France | 111244 | 7 | 30261 | 8 | 24908 | 8 | 63287 | 7 | 15119 | 8 | 15417 | 7 | 9533 | 8 | 22420 | 8 | 9383 | 9 | 30022 | 9 |
| Italy | 101875 | 8 | 32492 | 6 | 25239 | 7 | 53786 | 9 | 13852 | 9 | 12621 | 8 | 8678 | 9 | 20895 | 10 | 8822 | 12 | 32971 | 7 |
| Canada | 94366 | 9 | 31486 | 7 | 23877 | 9 | 45177 | 11 | 16672 | 6 | 11260 | 9 | 11166 | 7 | 17839 | 11 | 9440 | 8 | 30799 | 8 |
| Australia | 86629 | 10 | 29572 | 9 | 22457 | 11 | 38267 | 13 | 18039 | 5 | 8851 | 12 | 12022 | 4 | 14389 | 14 | 10547 | 7 | 28526 | 10 |
| Spain | 83918 | 11 | 24281 | 11 | 20520 | 12 | 42024 | 12 | 16055 | 7 | 9631 | 11 | 11563 | 6 | 15161 | 13 | 9359 | 10 | 23992 | 11 |
| Russia | 80025 | 12 | 8973 | 23 | 11135 | 17 | 59473 | 8 | 10016 | 12 | 7208 | 14 | 7085 | 10 | 21337 | 9 | 4033 | 17 | 10111 | 20 |
| South Korea | 79922 | 13 | 21920 | 14 | 18180 | 13 | 49287 | 10 | 5796 | 16 | 11041 | 10 | 4232 | 14 | 25882 | 7 | 5739 | 13 | 22815 | 12 |
| Brazil | 69935 | 14 | 22417 | 12 | 23639 | 10 | 29731 | 15 | 9673 | 13 | 6908 | 15 | 5817 | 13 | 10695 | 18 | 15080 | 3 | 21674 | 13 |
| Netherlands | 53779 | 15 | 21998 | 13 | 14560 | 14 | 21940 | 18 | 10227 | 11 | 5440 | 16 | 6703 | 11 | 7748 | 20 | 4852 | 15 | 21392 | 14 |
| Iran | 51156 | 16 | 12734 | 18 | 12042 | 15 | 31732 | 14 | 3535 | 30 | 5203 | 18 | 1827 | 35 | 16564 | 12 | 4297 | 16 | 15163 | 16 |
| Poland | 44503 | 17 | 9763 | 20 | 10123 | 18 | 27926 | 16 | 5051 | 20 | 5437 | 17 | 3337 | 20 | 11585 | 16 | 4908 | 14 | 10077 | 21 |
| Turkey | 43474 | 18 | 17376 | 15 | 8593 | 20 | 20099 | 21 | 5139 | 19 | 4218 | 22 | 3641 | 16 | 8617 | 19 | 3778 | 19 | 17560 | 15 |
| Switzerland | 41984 | 19 | 14928 | 16 | 11196 | 16 | 20775 | 19 | 5139 | 19 | 4349 | 21 | 3097 | 21 | 6456 | 23 | 3965 | 18 | 14309 | 17 |
| Sweden | 37821 | 20 | 13054 | 17 | 9841 | 19 | 18383 | 22 | 6055 | 14 | 4181 | 23 | 3958 | 15 | 6947 | 21 | 3699 | 21 | 12485 | 18 |

Thus, it is observed that relative research output ranks of different countries vary across different subject areas. Therefore, any assessment exercise that uses a subset of data (from selected subject areas) may produce outputs different from those obtained using whole data from all fields taken together. In order to look at all such pair-wise variations, we have also computed Spearman Rank correlations between rankings on all the subject areas, as explained below.

Table 4 present the matrix for SRCC values for ranks on different subject areas. It is observed that research output rank on SS&AH have the smallest correlation with other subject areas, indicating that different countries have significantly different amount of research output in this area. Similarly, among major areas, Physical Sciences and Health Sciences have SRCC value of 0.62, indicating different relative orders of countries in research output in these areas. SRCC values between ENG and Health Sciences is also low, again indicating differences in relative research outputs of different countries in these areas. Among specific subject areas, CS and ENG, and AGR, BIO & VET and MED, IMM & DEN subject areas have relatively higher pairwise rank correlations. The observations above, thus, clearly indicate that different countries have different strengths of research in different subject areas. Given that the relative research output volumes and ranks in different subject areas are not congruent, using a subset of research output data in an assessment exercise, may produce outcomes that are not only different from other subject areas but also from the overall research output data.

**Table 4: Spearman Rank Correlation Coefficients (SRCC) of different subject area rankings**

|  | All Fields | Health Sciences | Life Sciences | Physical Sciences | Social Science | CS | SS & AH | ENG | AGR, BIO & VET | MED, IMM & DEN |
|---|---|---|---|---|---|---|---|---|---|---|
| All Fields | 1 | 0.84 | 0.95 | 0.94 | 0.66 | 0.95 | 0.45 | 0.87 | 0.83 | 0.91 |
| Health Sciences | 0.84 | 1 | 0.91 | 0.62 | 0.64 | 0.76 | 0.42 | 0.54 | 0.77 | 0.97 |
| Life Sciences | 0.95 | 0.91 | 1 | 0.86 | 0.61 | 0.91 | 0.35 | 0.76 | 0.88 | 0.95 |
| Physical Sciences | 0.94 | 0.62 | 0.86 | 1 | 0.44 | 0.95 | 0.22 | 0.96 | 0.72 | 0.77 |
| Social Science | 0.66 | 0.64 | 0.61 | 0.44 | 1 | 0.58 | 0.95 | 0.28 | 0.61 | 0.61 |
| CS | 0.95 | 0.76 | 0.91 | 0.95 | 0.58 | 1 | 0.36 | 0.9 | 0.78 | 0.85 |
| SS & AH | 0.45 | 0.42 | 0.35 | 0.22 | 0.95 | 0.36 | 1 | 0.053 | 0.41 | 0.38 |
| ENG | 0.87 | 0.54 | 0.76 | 0.96 | 0.28 | 0.9 | 0.053 | 1 | 0.59 | 0.71 |
| AGR, BIO & VET | 0.83 | 0.77 | 0.88 | 0.72 | 0.61 | 0.78 | 0.41 | 0.59 | 1 | 0.8 |
| MED, IMM & DEN | 0.91 | 0.97 | 0.95 | 0.77 | 0.61 | 0.85 | 0.38 | 0.71 | 0.8 | 1 |

**Discussion**

The analytical results above highlight how the two exogenous factors (publication counting method and subject selection) can produce significantly different outcomes of research assessment exercises. It is observed that if the research output rankings are based on *whole* counting method, the NSF and DST-Elsevier reports obtain a very high rank correlation (with same country ranks from 1st to 19th place). However, the research output ranks of countries in NSF based on *fractional* counting are observed to be more different from DST- Elsevier report, despite the fact that they are very close in terms of subject area composition of data, and draw the data from the same database. Interestingly, a country like UK which is ranked at 3rd rank in terms of absolute research output moves to 6th rank if *fractional* counting is used. Similarly, India which is at 5th rank in terms of absolute research output moves to 3rd rank with use of *fractional* counting. There are several other examples illustrated in results. Thus, it can be said that use of *fractional* counting can produce significantly different outcomes as compared to use of *whole* counting. Similar observations were recorded in earlier studies by Gauffriau & Larsen (2005)[9] and Gauffriau et al. (2008)[10], where in it was concluded that publication counting methods are decisive for rankings based on publication and citation studies.

It is further observed in the analytical results that use of *fractional* counting impacts the rank of countries more which have very high or very low international collaboration. Countries like Switzerland get a reduction of as large as 83% in their publication score due to use of *fractional* counting. A country like UK, with significant publication volume, gets a significant reduction in publication score due to use of *fractional* counting, decreasing its research output rank from 3rd to 6th. On the contrary, countries that have lower international collaboration stand to gain in publication score, an example being India moving to 3rd rank from 5th rank on research output. Therefore, it is extremely important to understand the consequences of use of *fractional* counting in country-level assessment exercises. The use of *fractional* counting can artificially improve publication rank of a country, without any reflection on the country's overall research quality. Given that the previous studies (such as Glanzel (2001)[11] and Khor & Yu (2016)[12]) have shown that internationally collaborated research gets higher citations, use of *fractional* counting for country-level research assessment exercises need to be seen with this caution. In such cases, the important dimension of international collaboration in research gets masked, whereas use of the *whole* counting method would ease out this effect or situation. This consideration may have plausibly influenced OECD in using whole counting in its science and technology indicators scoreboard (see for example OECD and SCImago Research Group (CSIC), 2016[7]). However, it may also be noted that a study by Tarkhan-Mouravi (2019)[8] have shown that use of whole counting can inflate research outputs of some countries several times.

Therefore, as far as it stands for country-level assessment, it would be a better alternative to use *whole* counting, if rankings and assessment are expected to capture important dimension of international collaboration as well. In case, *fractional* counting method is used, the rankings for a country should be read in conjunction with the ICP. It is important to note here that while use of *fractional* counting may be a better reflection of actual contribution of an institution/ country in research, it also masks the important dimension of collaboration in research. Countries like Switzerland which have very high international collaboration, apparently due to its participation in several international projects, stand to lose significantly due to use of *fractional* counting. While higher collaborated output could be viewed in the perspective of lesser actual contribution from the country, yet the capacity of institutions in a country to engage in international collaboration is an important feature that benefits the country and its institutions. The *whole* counting method better captures the important dimension of international collaboration in research. Nevertheless, benefits of use of *whole* or *fractional* counting and to what extent they can more suitably represent the actual contribution in research, remains an important question worth exploring independently.

Another important thing to take into consideration here is that subject selection can also vary the research output ranks of countries. As observed in the results, research output rank of the same country varies significantly on data for different subject areas. For example, India is at 3rd rank in research output in Physical Sciences, 5th in Life Sciences, and 10th in Health Sciences and Social Science, both. Similarly, Russia is at 8th position in research output in Physical Sciences, 12th in Social Sciences, 17th in Life Sciences and 23rd in Health Sciences. Therefore, subject-specific assessments, though useful to understand the relative research strength of a country in a specific area, should not be taken as an overall evidence of research capability of a country. In fact, one may prefer to use an assessment based on wholistic data, comprising of all disciplines, for an overall picture. Subject-specific assessments can nevertheless indicate subject areas, in which a country should focus more in order to improve its overall position in global research landscape (Social Science is one such area in case of India). The point that could be understood from the observations here is that countries occupy different rank in terms of subject area, thus bundling a set of subject disciplines (e.g. NSF- S&E) vis-a –vis whole set of subject disciplines (e.g. NSF- S&T or DST- Elsevier) for a particular database (say Scopus) is bound to create variation in the overall rank of a country. The same will apply equally in case of data drawn from different databases (say Scopus, Web of Science, Dimensions). In other words, rank based on a mix of subject area (NSF S&E) and on overall subject areas (DST- Elsevier) for a country would vary when compared with comparators, both for the same or different databases used. While making these observations, it is also important to note that different databases use different subject classification schemes, including classification of articles into more than one subject area.

**Conclusion**

The paper analyses the impact of methodological approaches (mainly publication counting method and subject selection) on outcomes of research assessment exercises and provides meaningful conclusions.

*First*, use of reports based on *fractional* counting at country level should be read with other important dimensions like quality and international collaboration. A balanced understanding of research strength of a country needs inputs on several important dimensions, including citations and international collaboration, unfortunately use of *fractional* counting masks the important dimension of international collaboration.

*Secondly*, assessment exercises with a subset of research output data, including from selected disciplines, may have their own use cases, but may not be a true representation of overall research strength of a country. A wholistic assessment based on comprehensive data may be preferred for country-level studies.

This study thus contributes in enriching the methodological aspects that require careful considerations while undertaking studies based on research publications. It opens up for example, the current debate on the 'methodological dilemma' on *fractional* counting vis-à-vis *whole* counting (see for example, Egghe at al. 2000[13], Tarkhan-Mouravi, 2020[8]). The study also points out that any assessment exercise that uses a subset of data (from selected subject areas) may produce outputs different from those obtained using whole data from all fields taken together. We argue that scientometrics based studies require these types of insights to make the results more reliable and useful to the policy community at large.

Thus, the two-part study on factors that can affect the outcomes of research assessment reports helps in understanding impact of both, the endogenous factors (database-related) and exogenous factors (methodology-related). Both studies taken together present useful observations and implications for science administrators and policy makers. These studies, however, do not analyse other kinds of research outputs (such as patents) and the impact of number of journals from a country indexed in different databases, which would be an equally interesting thing to pursue as a future work.